\documentclass{eptcs}
\providecommand{\event}{TFPIE 2025} 

\usepackage[T1]{fontenc}
\usepackage[utf8]{inputenc}
\usepackage{iftex}
\usepackage{underscore}
\usepackage{hyperref}
\usepackage{fancyvrb}
\usepackage{xcolor}
\usepackage{graphicx}
\usepackage{subfigure}

\title{Teaching Introductory Functional Programming Using~Haskelite}

\newcommand{\titlerunning}{Teaching Introductory Functional Programming Using Haskelite}

\author{Pedro Vasconcelos 
  \institute{Department of Computer Science, Faculty of Science,
  University of Porto, Portugal}
  \email{pbvascon@fc.up.pt}  
}

\newcommand{\authorrunning}{Pedro Vasconcelos}

\newcommand{\comment}[1]{\textsf{\textcolor{gray}{#1}}}

\begin{document}
\maketitle

\begin{abstract}
  Learning functional programming requires learning a
  substitution-based computational model.
  While substitution should be a familiar concept
  from high-school algebra, students often have difficulty applying it
  to new settings, such as recursive definitions, algebraic
  data types and higher-order functions.  Step-by-step interpreters
  have been shown to help beginners by clarifying misconceptions and
  improving understanding.
  
  This paper reports on the experience of using a step-by-step tracing
  interpreter for a subset of Haskell while teaching an introductory
  functional programming course at the University of Porto.  We
  describe the use of the interpreter, present some 
  feedback obtained from students, reflect on the lessons learned
  and point directions for further work.
\end{abstract}


\section{Introduction}

Many Haskell textbooks explain the evaluation of pure functional
programs as a process of stepwise rewriting using
equations~\cite{birdwadler1988,thompson2011,bird2015,hutton2016}.
Traditionally these evaluations are done using paper and pencil.
However, we have noticed that students (and sometimes even
instructors) avoid performing these for anything but simple
computations because of the repetitive nature of the task.  Students
also make mistakes which may go unnoticed when doing these evaluations
on their own~\cite{Wilson2018}.  To help beginners avoid these
pitfalls and improve their understanding, we have
developed \emph{Haskelite}, a web-based interpreter for a subset of
Haskell that automatically constructs such rewrite
traces~\cite{Vasconcelos2023,VasconcelosMarques2024}.  The choice of a
web-based system eliminates the need to install special tools and
reduces the barrier to entry, which is particularly important with
beginners.

This paper reports on the experience of using this interpreter while
teaching an introductory functional programming course  at the
University of Porto. We describe how Haskelite was used for
demonstrations in lectures and also by the students in
practical classes and  self-study, present some qualitative
feedback obtained from students, reflect on the lessons learned, and
point directions for further work.


\section{Context}
\subsection{The course}
\emph{Programação Funcional} (CC1005) is an undergraduate course
offered by the Computer Science Department at the Faculty of Science
of the University of Porto.  The course is compulsory for first year
Computer Science majors, who have already taken an introductory
programming course in Python.  These make up the majority of our
students, but the course is also offered to students from other backgrounds,
who may not have had previous programming experience.

\begin{table}
  \begin{tabular}{c|p{0.8\textwidth}}
    \textbf{Lecture} & \hfill\textbf{Topics}\hfill\vspace{0ex}  \\ \hline
    1 & Expressions and values. Step-by-step evaluation.
        The Haskell interactive interpreter (GHCi).
        Some examples with arithmetic operations and functions
        from the Haskell Prelude. \\
    2 & Types as collections of values. Type assignment. Basic types
        for numbers, booleans and characters.
        Structured types for lists, tuples and functions. Polymorphic functions.
        Overloading and type classes.  \\
    3 & Defining functions using equations. Conditional expressions
        and guards. Pattern-matching with multiple equations.
        Patterns over tuples and lists. Case expressions
        and lambda expressions. \\
    4 & More about lists. De-sugaring lists into constructors.
        Arithmetic sequences. List comprehensions.
        Strings as a special case of lists. \\
    5 & Case study on list comprehensions: implementing the Caesar Cipher.\\
    6 & Recursive definitions.  Recursion over integers.
        Structural recursion over lists. Examples (\emph{sum},
        \emph{product}, \emph{length}).\\
    7 & Higher-order functions as abstractions of
        common recursion patterns.
        Some examples: \emph{map}, \emph{filter}, \emph{takeWhile},
        \emph{dropWhile}.
        Folding functions for lists: \emph{foldr} and \emph{foldl}.
        Folds as structural transformations of lists.\\
    8 & Infinite lists. Producing infinite lists using numeric sequences.
        Lazy evaluation for processing infinite lists
        (e.g.\@ mapping and filtering).
        Recursive definitions of infinite lists.    
        Useful functions from the Prelude (\emph{repeat}, \emph{cycle},
        \emph{iterate}). \\
    9 & Interactive programs. Actions versus values. The IO type constructor.
        Basic IO actions. Combining IO actions using \emph{do}-notation.
        Complete programs and batch compilation. \\
    10 & Case study on programming with IO: implementing John Conway's Game of Life.\\
    11 & Live coding example: determining the $N$ most common words in a text. \\
    12 & Live coding example: solving the 8 Queens Puzzle. \\
    13 & Defining new data types. Declaring enumeration types.
         Using constructors as expressions and in patterns. Constructors
         with arguments. Recursive data types. \\
    14 & Case study on recursive data types:
         implementing a tautology checker. \\
    15 & Binary search trees. \\
    16 & AVL binary search trees. \\
    17 & Proving properties about
         programs using equational reasoning and induction. \\ 
  \end{tabular}
  \caption{Lecture plan.} \label{tab:lectures}
\end{table}

The syllabus covers classic topics in functional programming
textbooks~\cite{hutton2016,bird2015,thompson2011,birdwadler1988}.
The weekly schedule consisted of two 1~h lectures and one 2~h
laboratory class.  Table~\ref{tab:lectures} summarizes the lecture
contents. The plan mostly follows Hutton's
textbook~\cite{hutton2016}, with one extra topic (AVL trees) from
Bird and Wadler's textbook~\cite{birdwadler1988}.

The experiment reported in this paper was conducted in the
Spring semester of 2024, where we used Haskelite alongside GHC/GHCi in
classes.  There were 143~students enrolled, of which 133 were Computer
Science majors, while the remaining were Mathematics or AI and Data
Science majors.

\subsection{The Haskelite interpreter}
\emph{Haskelite} is a step-by-step interpreter for a subset of Haskell that produces evaluation traces for a given expression.
The user interface presents an editor where users can provide their definitions and enter an expression to be evaluated.
Clicking the \emph{Evaluate} button switches to a view that allows users to control evaluation.
Currently, we provide the ability to step forward or backward through evaluation\footnote{Including non-terminating evaluations or ones that produce lazy structures (e.g.\@ infinite lists)}. 
This can be controlled using the on-screen buttons or the keyboard.
This evaluation view presents each step as an expression accompanied by a justification of the evaluation step, which can be either an equation from the program or a primitive operation.

Consider, for example, the
canonical definition of insertion into an ordered list.
Evaluating the expression \verb|insert 3 [1,2,4]| in Haskelite
produces the trace in Figure~\ref{fig:eval-insert}, where each
evaluation step is explained as either a primitive or the use of an
equation.\footnote{Ellipsis (\texttt{...}) mark omitted
  contexts e.g.\@ the current equation guard under evaluation.}
  
\begin{figure}[t]
\begin{verbatim}
insert x [] = [x]
insert x (y:ys) | x<=y = x:y:ys
                | otherwise = y:insert x ys
\end{verbatim}

\begin{Verbatim}[commandchars=\\\{\}]
   insert 3 [1, 2, 4] 
  \comment{\{ 3 <= 1 = False \}}
= .... False
  \comment{\{ insert x (y:ys) | otherwise = y:insert x ys \}}
= 1 : (insert 3 [2, 4])
  \comment{\{ 3 <= 2 = False \}}
= .... False
  \comment{\{ insert x (y:ys) | otherwise = y:insert x ys \}}
= 1 : (2 : (insert 3 [4]))
  \comment{\{ 3 <= 4 = True \}}
= .... True
  \comment{\{ insert x (y:ys) | x<=y = x:y:ys \}}
= 1 : (2 : (3  : (4 : [])))
  \comment{\{ final result \}}
= [1, 2, 3, 4]
\end{Verbatim}
\caption{Sample definition and evaluation trace.} \label{fig:eval-insert}
\end{figure}

Haskelite supports a subset of the Haskell~98 language
and one GHC extension:
\begin{itemize}
\item definitions with multiple equations and patterns;
\item lambda expressions, sections and partial application;
\item booleans, integers, characters and strings (i.e.\@ lists of characters);
\item tuples and lists;
\item boolean guards, if-then-else and case expressions;
\item \emph{where} and \emph{let} bindings;
\item new data type definitions and type aliases;
\item bang patterns\footnote{A GHC extension to force evaluation to weak head normal form.};
\item most list functions from the Prelude (see Appendix~\ref{app:prelude}).
\end{itemize}
The most salient omissions are:
\begin{itemize}
\item type classes; in particular, arithmetic operations work only on
  integers and (structural) comparisons are fully
  polymorphic\footnote{Hence can fail at runtime e.g.\@ when applied
    to function arguments.};
\item list comprehensions;
\item pattern bindings, i.e.\@ you can write \emph{let
    var = expression} but not \emph{let pattern = expression}
  for arbitrary patterns;
\item higher-kinded types, i.e.\@ all type variables
  must be of kind \texttt{*}; 
\item modules.
\end{itemize}
Despite the restrictions, Haskelite is suitable for running most
textbook programming examples, possibly with some adaptations.

An earlier version of Haskelite used a naive call-by-name reduction
strategy (i.e.\@ non-strict but not lazy) and supported a smaller
language subset~\cite{Vasconcelos2023}.  The version used for this
experiment is based on a lazy abstract
machine~\cite{VasconcelosMarques2024} for a pattern matching
calculus~\cite{kahl2004}.

The complete interpreter (including parser, type-checker, evaluator and
UI) is implemented in roughly 5000~lines of
\emph{Elm}\footnote{\url{https://elm-lang.org}} and compiled to
JavaScript; the resulting application runs directly from a web browser
and requires no installation. The source code is available at
\url{https://github.com/pbv/haskelite} and an online demo page can be
found at \url{https://pbv.github.io/haskelite}.

Elm was chosen as the implementation language for Haskelite because
(1) performance is not an overly critical concern for a teaching
interpreter, hence a high-level language is suitable; 
(2) a statically-typed functional language is a good match
for implementing a parser, type-checker and reduction machine;
(4) the Elm architecture is well-suited for
a self-contained interactive application; (5) the Elm compiler
produces small JavaScript
files, allowing the application to
start quickly.\footnote{The minimized JavaScript for the parser, type-checker, abstract machine and visualization is only 109~Kb.}

One drawback is that Elm is not as actively maintained as other more
mainstream functional languages; it also does not have as many
off-the-shelf libraries that we could reuse e.g., for parsing
Haskell source code.
Nonetheless, we believe the positives outweigh these
minor disadvantages.
  

\section{Haskelite in the Classroom}

We used Haskelite in conjunction with GHCi in lectures, alternating
between exposition and demonstration. GHCi was used when we wanted to
focus on types and values (i.e.\@ the \emph{denotations}) and
Haskelite when we wanted to focus on evaluation.  We also
engaged students with ``what if'' scenarios, e.g.\@ first showing an
evaluation to the entire class, then having them think about the effect of
some change, and using Haskelite interactively to check the answer. We
found that this helped to dispel misconceptions.

\paragraph{Simple functions.}
In Lecture~3 (defining functions using equations, guards, if-then-else
and patterns) we used Haskelite to highlight that equations and guards
are tried in order.  Consider, for example, a function to compute the
sign of an integer:
\begin{verbatim}
sign :: Int -> Int
sign x | x>0 = 1
       | x<0 = -1
       | otherwise = 0
\end{verbatim}
We can show the effect of omitting the \emph{otherwise} clause or
changing the order of the guards.  Alternatively, we can also re-write
the definition using a single equation and an if-then-else.

\paragraph{Recursive definitions.}
Recursion is introduced in Lecture~6.  The canonical initial example
is the factorial function, but more interesting examples concern
recursion over lists, because these are inductive values.
Functions \emph{sum} or \emph{product} are predefined in Haskelite,
but we  can redefine them for experimentation.
\begin{verbatim}
sum :: [Int] -> Int
sum [] = 0
sum (x:xs) = x + sum xs
\end{verbatim}
Evaluation with Haskelite can be used to show the roles of the base and
recursive cases (in particular, that the evaluation fails if we forget
the base case).  We can also show that changing the order of the
equations does not affect evaluation because the patterns are
non-overlapping.

We can then ask students to came up with an analogous
recursive definition for \emph{product}.  A typical mistake is to
define the base case incorrectly:
\begin{verbatim}
product [] = 0 -- wrong base case
\end{verbatim}
Figure~\ref{fig:sum-product} shows the Haskelite evaluation for the
correct \emph{sum} and incorrect \emph{product}, which should make it
clearer why the product of the empty list should be 1.  This can also
be used to point out significance of 0 of 1 as neutral elements
for~\texttt{+} and~\texttt{*}, respectively.

\begin{figure}
  \begin{minipage}{0.45\textwidth}
  \begin{Verbatim}[commandchars=\\\{\}]
  sum [1, 2, 3]   
  \comment{\{ sum (x:xs) = x + sum xs \}}
= 1 + (sum [2, 3])
  \comment{\{ sum (x:xs) = x + sum xs \}}
= 1 + (2 + (sum [3]))
  \comment{\{ sum (x:xs) = x + sum xs \}}
= 1 + (2 + (3 + (sum [])))
  \comment{\{ sum [] = 0 \}}
= 1 + (2 + (3 + 0))
  \comment{\{ 3 + 0 = 3 \}}
= 1 + (2 + 3)
  \comment{\{ 2 + 3 = 5 \}}
= 1 + 5
  \comment{\{ 1 + 5 = 6 \}}
= 6
\end{Verbatim}
\end{minipage}
\begin{minipage}{0.45\textwidth}
  \begin{Verbatim}[commandchars=\\\{\}]
  product [1, 2, 3]
  \comment{\{ product (x:xs) = x*product xs \}}
= 1 * (product [2, 3])
  \comment{\{ product (x:xs) = x*product xs \}}
= 1 * (2 * (product [3]))
  \comment{\{ product (x:xs) = x*product xs \}}
= 1 * (2 * (3 * (product [])))
  \comment{\{ product [] = 0 \}}
= 1 * (2 * (3 * 0))
  \comment{\{ 3 * 0 = 0 \}}
= 1 * (2 * 0)
  \comment{\{ 2 * 0 = 0 \}}
= 1 * 0
  \comment{\{ 1 * 0 = 0 \}}
= 0
\end{Verbatim}
\end{minipage}
\caption{Evaluation of a correct sum and incorrect product.}
\label{fig:sum-product}
\end{figure}

An alternative but less general solution is to define the base case
for the singleton list:
\begin{verbatim}
product [x] = x
product (x:xs) = x*product xs
\end{verbatim}
These equations do not define the product of the empty list, i.e.\@
they define a partial function -- and we should aim for total
functions whenever possible.
We can also use Haskelite to show that the equation for the
singleton list is redundant when we also have the case for
the empty list.

\paragraph{Eliminating concatenations.}
We revisit recursion in Lecture~17 when discussing equational
reasoning. For example, we derive an efficient list reverse from the
naive definition by eliminating concatenations.  Haskelite can then be
used to contrast the evaluation of the two definitions.
Figure~\ref{fig:reverse} shows the evaluation of fast reverse of a
4~character string.

This demonstration can be used to question the growth of the number of
steps with respect to the length $n$; it is then easy to argue that
the fast reverse requires one extra step for each character,
whereas the naive version requires $n$ steps to append to
the list --- hence the fast version has complexity $O(n)$ and the
naive has complexity $O(n^2)$.

\begin{figure}
\begin{verbatim}
fast_reverse xs = revAcc xs []
revAcc [] ys = ys
revAcc (x:xs) ys = revAcc xs (x:ys)
\end{verbatim}

  \begin{Verbatim}[commandchars=\\\{\}]
  fast_reverse "ABCD"   
\comment{\{ fast_reverse xs = revAcc xs [] \}}
= revAcc "ABCD" []    
\comment{\{ revAcc (x:xs) acc = revAcc xs (x:acc) \}}
= revAcc "BCD" ('A':[])
\comment{\{ revAcc (x:xs) acc = revAcc xs (x:acc) \}}
= revAcc "CD" ('B':('A':[]))
\comment{\{ revAcc (x:xs) acc = revAcc xs (x:acc) \}}
= revAcc "D" ('C':('B':('A':[])))
\comment{\{ revAcc (x:xs) acc = revAcc xs (x:acc) \}}
= revAcc [] ('D':('C':('B':('A':('H':[])))))
\comment{\{ revAcc [] acc = acc \}}
= 'D':('C':('B':('A':[])))
\comment{\{ final result \}}
= "DCBA"
  \end{Verbatim}
  \caption{Evaluation of fast reverse.}\label{fig:reverse}
\end{figure}

\begin{figure}
  \begin{minipage}{0.49\textwidth}
\begin{Verbatim}[commandchars=\\\{\}]
  append [1, 2, 3] [4, 5]
  \comment{\{ append xs ys = foldr (:) ys xs \}}
= foldr (:) [4, 5] [1, 2, 3]
  \comment{\{ foldr f z (x:xs) = f x (foldr f z xs) \}}
= 1 : (foldr (:) [4, 5] [2, 3])
  \comment{\{ foldr f z (x:xs) = f x (foldr f z xs) \}}
= 1 : (2 : (foldr (:) [4, 5] [3]))
  \comment{\{ foldr f z (x:xs) = f x (foldr f z xs) \}}
= 1 : (2 : (3 :
      (foldr (:) [4, 5] [])))
  \comment{\{ foldr f z [] = z \}}
= 1:(2:(3:[4, 5]))
  \comment{\{ final result \}}
= [1, 2, 3, 4, 5]
\end{Verbatim}
\end{minipage}
\begin{minipage}{0.5\textwidth}
\begin{Verbatim}[commandchars=\\\{\}]
  reverse [1, 2, 3]
  \comment{\{ reverse xs = foldl snoc [] xs \}}
= foldl snoc [] [1, 2, 3]
  \comment{\{ foldl f z (x:xs) = foldl f (f z x) xs \}}
= foldl snoc (snoc [] 1) [2, 3]
  \comment{\{ foldl f z (x:xs) = foldl f (f z x) xs \}}
= foldl snoc (snoc (snoc [] 1) 2) [3]
  \comment{\{ foldl f z (x:xs) = foldl f (f z x) xs \}}
= foldl snoc (snoc
                (snoc (snoc [] 1) 2)
                3) []
  \comment{\{ foldl f z [] = z \}}
= snoc (snoc (snoc [] 1) 2) 3
  \comment{\{ snoc ys x = x:ys \}}
= 3:(snoc (snoc [] 1) 2)
  \comment{\{ snoc ys x = x:ys \}}
= 3:(2:(snoc [] 1))
  \comment{\{ snoc ys x = x:ys \}}
= 3:(2:(1:[]))
  \comment{\{ final result \}}
= [3, 2, 1]
\end{Verbatim}
\end{minipage}
\caption{Evaluation of list append and reverse defined using folds.}\label{fig:fold-examples}
\end{figure}

\paragraph{Higher order functions.}

The fundamental higher-order functions \emph{map}, \emph{filter} and
\emph{foldl/foldr} are introduced in Lecture~7.  From prior years'
experience, we know that folds are significantly more difficult for
students than \emph{map} and \emph{filter}, hence we had intended
to use Haskelite for teaching them.

The pedagogical approach follows the one in Hutton's
textbook~\cite{hutton2016}, presenting first \emph{foldr} by as
abstracting a common recursive pattern of list processing functions
(of which the previous \emph{sum} and \emph{product} are instances).
We also emphasize the view of \emph{foldr f z} as a structural
transformation mapping the list constructor \texttt{(:)} to $f$ and
the empty list \texttt{[]} to $z$. Examples using small lists can be
demonstrated in Haskelite (see Figure~\ref{fig:fold-examples}).

\paragraph{Lazy evaluation.}
Lazy evaluation is mentioned informally at the start of the course but
only addressed explicitly in Lecture~8.  This lecture focuses on
defining and processing of infinite lists using enumerations (e.g.\@
\texttt{[1..]}), explicit cyclic definitions or Prelude functions
such as \emph{repeat} and \emph{cycle}.  Haskelite was used to help
build some operational intuitions for how infinite
structures can be  computationally effective.

A simple example is the following: write an expression to find all
positive integers whose square is less than 50.  A naive first attempt
might be
\begin{verbatim}
filter (\x -> x*x<50) [1..]
\end{verbatim}
which produces all the correct values, but does not terminate.
This can then be contrasted with the evaluation using \emph{takeWhile}:
\begin{verbatim}
takeWhile (\x -> x*x<50) [1..]
\end{verbatim}
Of course, such examples could also be tried in GHCi; however,
Haskelite makes the cause of non-termination apparent: the
recursive definition of \emph{filter} carries on indefinitely because
the condition $x\times x<50$ will never be true for $x>7$.

Haskelite can also be used to illustrate some more subtle aspects of
lazy evaluation, namely the use of sharing to construct cyclic
infinite structures.  Figure~\ref{fig:repeat-lists} shows the
evaluation of two definitions of the $repeat$ function that
construct an infinite list of repeated values.  While the definitions
are denotationally equivalent, they result in very different
operational behaviors: $repeat'$ constructs a cyclic list in a
single step while $repeat$ expands the list one element at a
time.  Note that the outermost evaluation forces the expansion of
constructor arguments; however, this would not terminate in the case of
cyclic structures, so it is suspended with a ``\emph{continue?}''
message.

\begin{figure}
  \begin{minipage}{0.41\textwidth}
\begin{verbatim}
repeat x = x : repeat x
\end{verbatim}
    \vfill
  \begin{Verbatim}[commandchars=\\\{\}]
  repeat 1
\comment{\{ repeat x = x:repeat x \}}
= 1:(repeat 1)
\comment{\{ repeat x = x:repeat x \}}
= 1:(1:(repeat 1))
\comment{\{ repeat x = x:repeat x \}}
= 1:(1:(1:(repeat 1)))
\comment{\{ repeat x = x:repeat x \}}
= 1:(1:(1:(1:(repeat 1))))
\end{Verbatim}
\end{minipage}
\begin{minipage}{0.50\textwidth}
\begin{verbatim}
repeat' x = xs
   where xs = x:xs
\end{verbatim}
    \vfill    
    \begin{Verbatim}[commandchars=\\\{\}]
 repeat' 1
\comment{\{ repeat' x = xs \}}
= xs$0
\comment{\{ xs = x:xs \}}
= 1:xs$0
\comment{\{ continue? \}}
= 1:(1:(1:(1:(1:(1:(1:(1:(1:(1:xs\$0)))))))))
\end{Verbatim}
\vfill
\vspace{1ex}
\end{minipage}
\caption{Evaluations of two definitions of repeat.}\label{fig:repeat-lists}
\end{figure}

\begin{figure}
\begin{verbatim}
data BST a = Leaf | Node a (BST a) (BST a)

insert :: a -> BST a -> BST a -- no Ord typeclass
insert x Leaf = Node x Leaf Leaf
insert x (Node y lt rt)
         | x<=y      = Node y (insert x lt) rt
         | otherwise = Node y lt (insert x rt)

makeTree :: [a] -> BST a -- no Ord typeclass
makeTree [] = Leaf
makeTree (x:xs) = insert x (makeTree xs)
\end{verbatim}

  \begin{Verbatim}[commandchars=\\\{\}]
  makeTree [1, 3, 2]
  \comment{\{ makeTree (x:xs) = insert x (makeTree xs) \}}
= insert 1 (makeTree [3, 2])
  \comment{\{ makeTree (x:xs) = insert x (makeTree xs) \}}
= .... insert 3 (makeTree [2])
  \comment{\{ makeTree (x:xs) = insert x (makeTree xs) \}}
= ........ insert 2 (makeTree [])
  \comment{\{ makeTree [] = Leaf \}}
= ............ Leaf
  \comment{\{ insert x Leaf = Node x Leaf Leaf \}}
= ........ Node 2 Leaf Leaf
  \comment{\{ 3 <= 2 = False \}}
= ........ False
= .... Node 2 Leaf (insert 3 Leaf)
  \comment{\{ 1 <= 2 = True \}}
= .... True
  \comment{\{ insert x (Node y lt rt) | x<=y = Node y (insert x lt) rt \}}
= Node 2 (insert 1 Leaf) (insert 3 Leaf)
  \comment{\{ insert x Leaf = Node x Leaf Leaf \}}
= Node 2 (Node 1 Leaf Leaf) (insert 3 Leaf)
  \comment{\{ insert x Leaf = Node x Leaf Leaf \}}
= Node 2 (Node 1 Leaf Leaf) (Node 3 Leaf Leaf)
  \comment{\{ final result \}}
= Node 2 (Node 1 Leaf Leaf) (Node 3 Leaf Leaf)
\end{Verbatim}
\caption{Building a binary search tree from a list.}\label{fig:insert-bst}
\end{figure}

\paragraph{Evaluating co-recursive infinite lists.}
We also use Haskelite for evaluating co-recursive definitions of
the infinite lists, e.g.\@ the well-known one-line definition of the
Fibonacci numbers:
\begin{verbatim}
fibs = 0 : 1 : zipWith (+) fibs (tail fibs)
\end{verbatim}
Because of lazy evaluation, obtaining the $n$-th element of
\emph{fibs} will take $O(n)$ steps.\footnote{Assuming a constant cost addition e.g.\@ over fixed precision integers.}
However, note that pretty-printing will expand shared
sub-expressions; thus it
may appear as if computations were duplicated until
they are simultaneously replaced by the weak normal form.
This choice is conscious:  
we want to show traces as term reductions for simplicity,
whereas the implementation is actually
performing graph reductions.

\paragraph{Binary search trees.}
We introduce algebraic data types in Lecture~13 and~14. Binary search
trees are covered in Lectures~15 and~16.  We can demonstrate standard
search tree algorithms using Haskelite. Figure~\ref{fig:insert-bst}
shows the evaluation of the construction a binary search tree from a
list of integers.\footnote{The type signatures omit the \emph{Ord}
  type constraint because Haskelite does not support type classes.}
The resulting tree respects the ordering property (i.e.\@ it is a
search tree) but is not always balanced; we can show that by inserting
the same elements in a different order.
This motivates the introduction of AVL trees~\cite{birdwadler1988}
which ensure both ordering and balance.



\section{Student Feedback}\label{sec:feedback}

\subsection{Feedback during classes}

\begin{table}
  \begin{tabular}{cp{32em}c}
    \textbf{Student} & \hfill\textbf{Comments}\hfill\vspace{0ex} & \textbf{Code}  \\ \hline
    1  & ``Helped me understand foldr and the use of anonymous functions'' & +\\
    2 &  ``Helped me understand recursion (\ldots). Step-by-step evaluation
        in Haskelite (\ldots) made recursion more understandable.''& + \\
     & ``The UI is is not very understandable at first,
        only with the help of some colleagues did I manage to
        run my program.'' & - \\
    3 &  ``Haskelite helped me better understand foldl and foldr, which initially I thought were very similar.'' & + \\
    4 & ``Haskelite was useful for solving some exercises on lists and recursive functions. A specific case was the function map (\ldots).'' & + \\
      & ``The visualization
        was fundamental to consolidate my understanding of recursion
        and higher-order functions.'' & + \\
    5 & ``I could easily understand (\ldots) higher-order functions,
        for example foldr and foldl (\ldots). I was able to see that foldr is efficient with infinite lists and foldl with finite lists'' & + \\
    6 & ``Haskelite helped me understand an exercise on recursively summing values in a binary tree (\ldots).'' & + \\
    & ``It was also helpful in understanding recursive functions from
        the previous worksheet.'' & + \\
     & ``List comprehensions (\ldots) help understand code more easily,
        so it would be helpful if Haskelite supported them.''& - \\
    7 & ``Helped me understand insertion into a binary search tree, and then
        helped solve exercises from worksheet 7 [on binary search trees].'' & + \\
    8 & ``Haskelite helped me understand folding much better,
        with which I was having trouble with in the worksheets (\ldots).'' & +\\
       & ``Also was useful in solving the worksheet on binary search trees.'' & + \\
    9 & ``I found Haskelite extremely useful for understanding
        how to use foldr and foldl for aggregating lists.'' & + \\
    10 & ``Haskelite helped me understand simple recursive functions
         such as intersperse [from Data.List].'' & + \\
    11 & ``Haskelite was helpful to debug some programs (\ldots)
         was well as to understanding recursion, foldr, foldl and trees.'' & + \\
    12 & ``When defining the takeWhile function (\ldots) the [syntax] error message is not very clear.'' & - \\
    13 & ``When evaluating an expression involving combinations
        of higher-order functions such as filter and foldl,
        it became difficult to the step-by-step evaluation.'' & - \\
    14 & ``Haskelite doesn't accept code that GHC accepts [because of list comprehensions].'' & - \\
    & ``Layout [of the evaluation steps] is unnecessary and makes the expressions more complex than necessary.'' & - \\
    \hline
  \end{tabular}
  \medskip
  \caption{Summary of students comments and coding as
    positive or negative feelings towards Haskelite.}\label{tab:feedback}
\end{table}

We asked students to provide feedback on the use of Haskelite
during the semester.  This was done using the course's \emph{Moodle}
forum through two pinned discussion
topics, one for examples in which Haskelite had been helpful
and another for reporting difficulties using it.

Only 14~students provided feedback; this low participation (roughly
10\% of enrolled students) is also typical of the standard
end-of-semester evaluation surveys (in the academic year
2023/2024 there were only 15 responses).  It is also not possible to
fully determine how representative the sample is in terms of gender or
race (especially since students are not required to provide this
information when they enroll).  Nonetheless, using first names as a
proxy for gender classification, we had 3 female and 11 male
respondents, which is a similar ratio to that of the enrolled students
(26\% female).

The students comments (translated from Portuguese) are summarized in
Table~\ref{tab:feedback}, together with a lightweight coding performed
by the author into ``positive'' or ``negative'' statements towards
Haskelite.  The following are some reflections on the feedback
provided and lessons we have learned from it.
\begin{itemize}
\item A large majority of statements\footnote{Some
  students expressed more than one statement, hence the number of
  statements is higher than the number of respondents.} are positive (14 vs.\@ 6).
\item 11 students reported Haskelite helping
  with understanding recursion and/or higher-order functions.
\item 6 students mentioned help in understanding
  specifically the \emph{foldl} and \emph{foldr} functions.
  This is consistent with our perception that beginners
  typically find these more difficult than (say)
  \emph{map} and \emph{filter}.
\item 4 students mentioned Haskelite helping in understanding
  functions over tree-like structures.
\item Student 5 stated that Haskelite helped in understanding the processing
  of finite vs.\@ infinite lists. The remark on ``efficiency of foldr''
  is likely to refer that, with infinite lists, \emph{foldr} can be
  productive, whereas \emph{foldl} will never terminate.
\item 2 students wished for Haskelite to support list comprehensions;
  we discuss this limitation in Section~\ref{sec:conclusion}.
\item Student 12 expressed difficulties with the syntax
  error messages. This is most likely due to the parser rejecting programs
  using unsupported features (e.g.\@ list comprehensions or
  pattern bindings).
\item Student 13 reported difficulty understanding 
  the evaluation of compositions of higher-order functions.
\end{itemize}

\subsection{Anonymous feedback}
We have also conducted an anonymous questionnaire at the end of the
semester to gather feedback on the use of Haskelite in lectures and
to find out how students used the system independently.
Unfortunately, only 6~students responded to the questionnaire,
so the results are not statistically significant.
Nonetheless, we present them with the caveat that they are might not
be representative of the student population.
The research questions we are trying to answer are:
\begin{description}
\item[RQ1] Did students find the use of Haskelite in lecture demonstrations useful?
\item[RQ2] Did students find the use of Haskelite useful for self study?
\item[RQ3] [Sub-question] For which topics (if any) did students find Haskelite more useful?
\end{description}

For assessing RQ2 and sub-question RQ3, we asked students to report
how they used Haskelite in practical
classes and for autonomous study; the results are summarized in
Table~\ref{tab:practicals} (note that participants could select multiple
options if desired). 
\begin{table}
\begin{center}
\begin{tabular}{p{27em}l}
  \multicolumn{2}{l}{\textbf{Did you use Haskelite in practical classes?}} \\ \hline
    Yes, to help understand my solution attempts & 3 (50\%) \\
  Yes, to review and understand concepts & 2 (33\%) \\
  Yes, to debug problems & 4 (66,7\%) \\
  No & 2 (33,3\%)
\end{tabular} \\
\begin{tabular}{p{28em}l}
  \multicolumn{2}{l}{\textbf{If you answered ``No'' above, why did you not use Haskelite?}} \\ \hline
  I didn't find it necessary & 0 (0\%)\\
  I preferred to use GHC/GHCi & 1 (33,3\%) \\
  Because Haskelite does not support full Haskell & 1 (33,3\%) \\
  I tried to use it but couldn't figure out an error message & 2 (66,6\%) \\
  I used it for initial programs but stopped using for more complex ones & 0 (0\%)
\end{tabular}\\
  \begin{tabular}{p{27em}l}
    \multicolumn{2}{l}{\textbf{For which of the following topics
    did you use Haskelite in your self study?}} \\ \hline
  Simple functions on lists  & 0 (0\%)\\
  Recursive functions on lists & 3 (50\%)\\
  Simple functions on integers & 0 (0\%)\\
  Recursive functions on integers & 0 (0\%)\\
  Definitions using map or filter & 3 (50\%)\\
  Definitions using foldr or foldl & 5 (83,3\%)\\
  Definitions using other higher-order functions &  2 (33,3\%)\\
  Definitions over other recursive data types & 3 (50\%)
  \end{tabular}
\end{center}
  \caption{Usage of Haskelite in practicals and for self study.}
  \label{tab:practicals}
\end{table}

\begin{figure}[t]
  \begin{center}
    \includegraphics[width=0.5\textwidth]{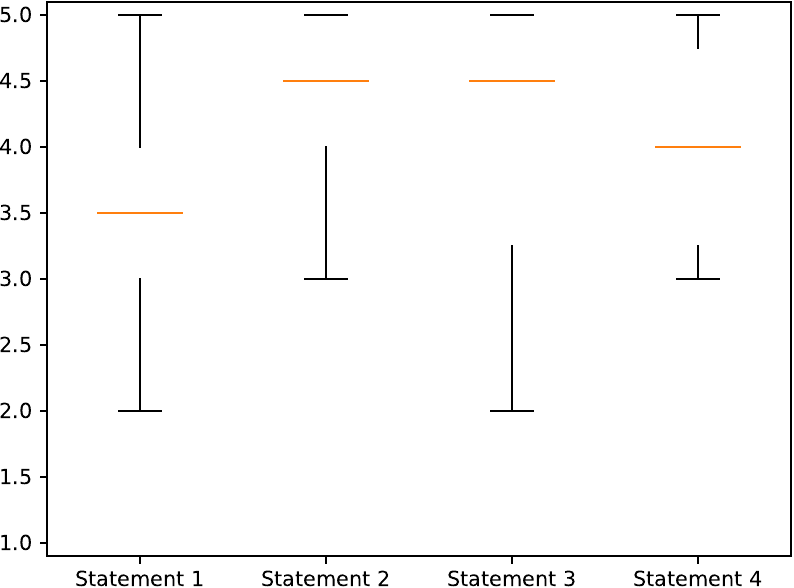}
  \end{center}
  \caption{Responses on the helpfulness of using Haskelite in lectures on
    pattern-matching, recursion, higher-order functions
    and binary search trees.
    Results are in a 5-point Likert scale (1=completely disagree, 5=complete agree).
    The red bars represent medians and the whiskers represent the minimum and
maximum values.
  } \label{fig:boxplot-lectures}
\end{figure}

For assessing RQ1 and sub-question RQ3 we asked students
to express their agreement with the
following four statements in a 5-point Likert
scale (from 1 meaning ``totally disagree'' to 5 meaning ``completely agree'').
\begin{description}
\item[Statement 1] ``The use of Haskelite in lectures helped me understand 
  pattern matching.''
\item[Statement 2] ``The use of Haskelite in lectures helped me understand
  recursion.''
\item[Statement 3] ``The use of Haskelite in lectures helped me understand
  higher-order functions.''
\item[Statement 4] ``The use of Haskelite in lectures helped me understand
  binary search trees.''
\end{description}
The results are shown in Figure~\ref{fig:boxplot-lectures}.    We note that the median of agreement with
statements~2 and~3 is higher than for the other two statements, which
is consistent with the comments obtained from Moodle that Haskelite
was particularly useful for understanding recursion and higher-order
functions.  However, the wide amplitude of answers indicates that the
opinions where not consensual.

Because all four statements in the anonymous feedback are positively
framed there a possibility of acquiescence bias in the responses. We
can attempt to estimate this bias using the coding in
Table~\ref{tab:feedback}: the ratio of positive statements towards Haskelite
is $14/(14+6) = 70\%$. In our Likert scale this would correspond to a
value of $3.8$, which this is slightly lower than the medians of replies
in Figure~\ref{fig:boxplot-lectures} but still positive.

\subsection{Comparison of success rates with previous years}

\begin{table}
  \begin{center}
  \begin{tabular}{l|c|c|c}
     & \textbf{2023/2024} & 2022/2023 & 2021/2022 \\ \hline
    Enrolled & 143        & 131        & 128 \\
    Not assessed & 31 (22\%)     & 31 (24\%) & 51 (40\%) \\
    Approved  & 88 (61\%)  & 39 (37\%) & 53 (41\%) \\
    Failed   & 24 (17\%)  & 51 (39\%) & 24 (19\%)   
  \end{tabular}
  \end{center}
  \caption{Success rates in different editions of the course.}\label{tab:success}
\end{table}

Table~\ref{tab:success} presents a comparison of the success rates of
students in the semester reported in this experiment (2023/2024)
against those of the previous two editions where the use of Haskelite
was not encouraged.  The line labeled ``Not assessed'' refers to
students who did not attend the final exam, either by choice or
because did not meet the necessary attendance criteria.  While the
total number of enrolled students has increased slightly, we observe a
substantially higher approval rate in the last edition.  Without
further study we cannot attribute this improvement to the use of
Haskelite alone, but it is nonetheless an encouraging result.


\section{Lessons Learned}\label{sec:lessons}

\paragraph{Choosing adequate examples.}
Evaluation traces for even innocent looking expressions can easily grow
and become unwieldy.  When choosing examples for lectures, we found it
best to pick simple examples:
\begin{itemize}
\item use small numbers and lists to limit the total number of
  evaluation steps;
\item use fully evaluated arguments when exemplifying recursion,
  e.g.\@  \texttt{[1,2,3,4,5]} instead of \texttt{[1..5]}
  (which is de-sugared into an application of \emph{enumFromTo});
\item use binary operators or named functions rather than
  lambda-expressions as arguments for higher-order
  functions;
\item be particularly careful with the size of arguments for examples
  of function composition, e.g.\@ the example for building a binary
  search tree from a list in Figure~\ref{fig:insert-bst}; this 
  relates to the comment by Student~13 in Table~\ref{tab:feedback};
\end{itemize}
  
\paragraph{The importance of error messages.}
Many students expressed difficulty understanding error messages,
particularly syntax errors caused by attempting to use unsupported
language features, e.g.\@ list comprehensions, pattern bindings,
floating point operations, etc. For example, given a definition
\begin{verbatim}
squares n = [x*x | x <- [1..n]]
\end{verbatim}
Haskelite would initially report an unhelpful error message:
\begin{verbatim}
line 1,col 9: expecting operator
\end{verbatim}
We have since improved the interpreter to parse more
expressions and reject the program during type checking with an
explicit error.  For the above example, the revised message is: 
\begin{verbatim}
definition of squares, expression [n*n | n<-[1..n]]:
list comprehensions are not supported
\end{verbatim}

\paragraph{Avoiding an excessively operational mindset.}
Although Haskelite was never intended as a debugger, we have witnessed
in classes that some students attempt to use it as such.  The
anonymous feedback presented in Table~\ref{tab:practicals} also
confirms this practice.  We conjecture that this ``debugging mindset''
may come from previous experience e.g.\@ with \emph{Python Tutor}, a
step-by-step interpreter for Python~\cite{pythontutor}.
Paradoxically, this means that it might actually be preferable that
Haskelite does \emph{not} support full Haskell, so that students are
discouraged from trying to understand larger programs using only the
operational view!

The tension between the operational and denotational view of programs
when teaching programming is well-known\footnote{For example: Abelson
  et al.\@ calls it ``declarative and imperative
  descriptions''~\cite{abelson1996}, while Bird and Wadler call it
  ``specifications and implementations''~\cite{birdwadler1988}.}, so
it should be no surprise to see it re-appear here.  Our attempt to
mitigate this is to teach functional decomposition by example: we used
some lectures for case studies and live coding sessions solving
classic problems (see Table~\ref{tab:lectures}).  We used the white
board for interactively sketching, a text editor for programming and
GHCi for testing, but \emph{not} Haskelite.  Students were also
encouraged to solve some take-home exercises that required separately
developing smaller functions and combining them into a larger problem.


\section{Related Work}

There is an long tradition in the functional programming community of
teaching languages and tools based on dialects of the Scheme language.
The \emph{DrRacket} programming environment includes an algebraic
stepper for the teaching subsets of Racket~\cite{htdp}.  Like Haskelite, the
DrRacket stepper presents evaluation as a sequence of algebraic
rewriting steps over the source program and is based on a high-level
reduction semantics~\cite{Clements2001}.  However, unlike
Haskelite, the supported languages employ a call-by-value semantics
rather than lazy evaluation, and do not promote the programming style
using equations and pattern matching that we aim to teach.

Tunner Wilson et al.~\cite{Wilson2018} performed an evaluation of the
students ability to trace recursive programs and found that students
performed better with the Scheme algebraic stepper than with a
traditional imperative model in Python, particularly with recursion
over tree-shaped data. This is consistent with our own findings
reported in this paper.

In the context of the ML family of languages, Furukawa et
al.~\cite{Furukawa2019} presented an algebraic stepper for a subset
of OCaml that includes exceptions and imperative features as well as
the functional core.  Like Haskelite, this stepper was used for
teaching an introduction functional programming course, and
the authors report that it was useful for improving
students understanding of recursive functions and data.
Asai and Akiyama~\cite{Asai2025} have extended this stepper
to allow limited support for OCaml modules.

Olmer et al.~\cite{olmerevaluating2014} developed a tutoring
environment for a subset of Haskell that includes a step-by-step
evaluator.  The evaluator is based on a general tool for defining
rewriting systems and supports different strategies (innermost or
outermost); it can also be used to check student's traces against an
expected strategy.  However, unlike Haskelite, it does not implement
lazy evaluation and does not support equation guards. Also unlike
Haskelite, it does not allow students to define their own functions.


\section{Conclusion and Further Work}
\label{sec:conclusion}
We found that using Haskelite in classes alongside GHCi was helpful in
teaching principles of functional programming, particularly for
explaining recursion and higher-order functions.  The feedback
obtained from students (albeit with a small number of participants)
also supports this.  Several directions of future extensions to this
work are possible.

\paragraph{Evaluating list comprehensions.}
We do not currently support evaluation of list comprehensions; these
could be added by translation into higher-order functions
(Section~3.11 of the Haskell 2010 report~\cite{haskell2010report}) but
such translation is likely to confuse students, since higher-order
functions are introduced after list comprehensions in our course
(following~\cite{hutton2016}).

Alternatively, we could implement the evaluation directly by
rewriting list comprehensions instead of translation. While
this would certainly be possible, none of the functional programming
textbooks that we are aware of takes this approach; this would thus
requires extra explanation to the students.  Because of these
challenges, we leave this topic for future reflections.

\paragraph{More flexible evaluation strategy.}
Currently, there is no mechanism for skipping evaluation steps; this
would be useful for focusing on some functions while omitting others.
For example, when demonstrating the classical list Quicksort example, we
might want to skip over the evaluation of list append.  This would be
easily done in a language with strict semantics because the lexical scopes
matches the order of evaluation; in a lazy semantics, the evaluation of
nested sorting occurs interleaved with appends, so this is not so
straightforward.

One method for skipping a function could be to treat it as ``hyper-strict''
(i.e.\@ forcing both the argument and result to full normal form) and
eliding all its evaluation steps.  This means that Haskelite would no
longer follow the Haskell semantics exactly, but (subject to
termination) should yield identical results.
We leave this as a topic for further work.

\paragraph{Using Haskelite in an advanced course.}
Haskelite supports some features that are we are not using in our
introductory course:
\begin{itemize}
\item \emph{Bang patterns} allow forcing the evaluation of function arguments,
  this can be used to contrast strict vs.\@ lazy implementations
  of similar functions (e.g.\@ $\mathit{foldl}$ and $\mathit{foldl'}$);
\item \emph{Non-regular data types and polymorphic recursive
  functions} allow for some more advanced data
  structures e.g.\@ based on numeric
  representations~\cite{okasaki1999}.
\end{itemize}
It would be interesting to explore the use of Haskelite 
in a more advanced functional programming course e.g.\@ on
purely-functional data structures.


\appendix
\section{Prelude functions}\label{app:prelude}

This appendix lists all Prelude functions predefined in Haskelite;
users may re-define any of these functions if they wish.

\begin{verbatim}
even, odd :: Int -> Bool
max, min :: a -> a -> a        -- overly polymorphic
compare :: a -> a -> Ordering  -- overly polymorphic
fst :: (a,b) -> a
snd :: (a,b) -> b
(&&), (||) :: Bool -> Bool -> Bool
head :: [a] -> a
tail :: [a] -> [a]
(++) :: [a] -> [a] -> [a]
(!!) :: [a] -> Int -> a 
length :: [a] -> Int
reverse :: [a] -> [a]          -- defined using an accumulator
init :: [a] -> [a]
last :: [a] -> a  
sum, product :: [Int] -> Int
and, or :: [Bool] -> Bool
take, drop :: Int -> [a] -> [a]
maximum, minimum :: [a] -> a   -- overly polymorphic
concat :: [[a]] -> [a]
repeat :: a -> [a]
replicate :: Int -> a -> [a]  
cycle :: [a] -> [a]
iterate :: (a -> a) -> a -> [a]
map :: (a -> b) -> [a] -> [b]
filter :: (a -> Bool) -> [a] -> [a]
foldr :: (a -> b -> b) -> b -> [a] -> b
foldl :: (a -> b -> a) -> a -> [b] -> a
foldl' :: (a -> b -> a) -> a -> [b] -> a   -- strict version of foldl 
(.) :: (b -> c) -> (a -> b) -> a -> c
($), ($!) :: (a -> b) -> a -> b            -- lazy and strict application
zip :: [a] -> [b] -> [(a,b)]
zipWith :: (a -> b -> c) -> [a] -> [b] -> [c]
takeWhile, dropWhile :: (a -> Bool) -> [a] -> [a]
any, all :: (a -> Bool) -> [a] -> Bool
lookup :: a -> [(a,b)] -> Maybe b   -- overly polymorphic
chr :: Int -> Char
ord :: Char -> Int
isAlpha, isDigit, isAlphaNum, isUpper, isLower :: Char -> Bool
show :: Int -> String   -- only for integers
-- list enumerations restricted to integers
enumFrom :: Int -> [Int]
enumFromTo :: Int -> Int -> [Int]
enumFromThen :: Int -> Int -> [Int]
enumFromThenTo :: Int -> Int -> Int -> [Int]
\end{verbatim}


\section*{Acknowledgments}
The author would like to thank Rodrigo Marques, Rita Ribeiro
and the anonymous reviewers
for helpful comments and suggestions on earlier drafts of this paper.

This work was partially supported by the Artificial Intelligence
and Computer Science Laboratory (LIACC) under the base research grant
FCT UIDB/00027/2020
of the FCT (Funda\c{c}\~{a}o para a Ci\^{e}ncia e
Tecnologia --- \texttt{https://www.fct.pt}) funded by national funds
through the FCT/MCTES (PIDDAC).


\bibliography{bibliography}

\end{document}